\documentclass[twocol]{ametsocV5_navid}

\usepackage{amsmath,amsfonts,amssymb,bm}
\usepackage{mathptmx}
\usepackage{newtxtext}
\usepackage{newtxmath}
\usepackage{wasysym}

\PassOptionsToPackage{hyphens}{url}\usepackage{hyperref}
\hypersetup{
	breaklinks,
	colorlinks=true,
	linkcolor=blue,
  urlcolor=Purple,
	citecolor=Purple,
 }

\renewcommand{\vec}{\boldsymbol}
\renewcommand{\d}[2]{\frac{\mathrm{d} #1}{\mathrm{d} #2}} 
 
\renewcommand{\deg}{^\circ}
\newcommand{\kv}{\hat{\vec{k}}}

\title{How winds and ocean currents influence the drift of floating objects}

\authors{Till J. W. Wagner\correspondingauthor{Till Wagner, till.wagner@wisc.edu}}
\affiliation{University of Wisconsin--Madison, Madison, WI \\ University of North Carolina Wilmington, Wilmington, NC}

\extraauthor{Ian Eisenman}
\extraaffil{\small Scripps Institution of Oceanography, University of California San Diego, La Jolla, CA}

\extraauthor{Amanda M. Ceroli}
\extraaffil{\small University of North Carolina Wilmington, Wilmington, NC}

\extraauthor{Navid C. Constantinou}
\extraaffil{\small Australian National University, Australian Capital Territory, Australia}

\abstract{Arctic icebergs, unconstrained sea ice floes, oil slicks, mangrove drifters, lost cargo containers,  and other flotsam are known to move at 2-4\% of the prevailing wind velocity relative to the water, despite vast differences in the material properties, shapes, and sizes of objects. Here, we revisit the roles of density, aspect ratio, and skin and form drag in determining how an object is driven by winds and water currents. Idealized theoretical considerations show that although substantial differences exist for end members of the parameter space (either very thin or thick and very light or dense objects), most realistic cases of floating objects drift at $\approx 3$\% of the free-stream wind velocity (measured outside an object's surface boundary layer). This relationship, known as a long-standing rule of thumb for the drift of various types of floating objects, arises from the square root of the ratio of the density of air to that of water. We support our theoretical findings with flume experiments using floating objects with a range of densities and shapes.}

\begin{document}

\maketitle

\section{Introduction}

The drift of a wide range of floating objects in geophysical settings follows separate empirical rules-of-thumb that all predict motion at 2-4\% of the wind velocity, relative to the velocity of the water. For example, observations consistently show that freely drifting Arctic icebergs travel at the water velocity plus 1.6-1.8\% of the near-surface wind velocity \citep{smith1983,garrett1985,bigg1997}. Unconstrained sea ice floes, similarly, have long been observed to typically drift at 2-2.5\% of the wind velocity relative to the ocean current \citep{nansen1902oceanography,zubov1945ice,browne1958,thorndike1982}.  Comparable drift behavior has separately been found for mangrove propagules in laboratory flume experiments \citep{vanderstocken2015}. The drift of human survivors of ship wreckages as well as that of man-made objects such as life rafts, cargo containers, and other types of flotsam have been the subject of an extensive body of research which is often referred to as ``search and rescue''  literature and which reports similar drift behavior \citep[e.g.,][]{allen1999review,daniel2002drift,allen2005leeway,breivik2011wind,breivik2012leeway,rohrs2012observation,breivik2013advances,nesterov2018consideration,sutherland2020evaluating}. Finally, oil slicks have also been observed to drift at approximately 3\% of the wind velocity relative to the ocean surface \citep{stolzenbach1977review,rasmussen1985}. Despite the wide range of shapes, materials, and aspect ratios, all of these types of objects demonstrate similar drift behavior.

Here, we consider the origin of these approximately equivalent wind sensitivities.  We begin with an idealized theoretical analysis of the momentum balance, focusing on the water and air drag forces on a partially submerged rectangular object. This general approach, which results in the drift velocity of a floating object being given as that of the near-surface ocean current velocity plus a small percentage of the ambient wind velocity (the leeway factor), has been referred to as the ``leeway modeling approach'' \citep{olascoaga2020observation}. Our idealized theoretical considerations (Sections 2 and 3) thus share a number of underlying assumptions with some previous work on leeway drift modeling \citep[e.g.,][]{daniel2002drift,nesterov2018consideration}, as discussed below. The results provide a physically intuitive explanation for the 2-4\% drift laws. Our framework also allows us to assess under what conditions the sensitivity to wind forcing deviates substantially from this value. Finally, in Section 4 we present the results of flume tank experiments, which support the theoretical results. 

Note that the idealized considerations presented here are not intended for use in operational drift forecasting, but rather to gain fundamental physical understanding of how wind and ocean current drags balance and influence the drift of floating objects. To accurately predict the actual trajectory of drifters, a framework is needed that accounts for a number of other pertinent forces, including inertial effects, lift and added mass forces, and the Coriolis force. Such a comprehensive account is given by the Maxey--Riley set for surface ocean inertial particles, and a series of recent studies have made substantial progress toward accurately describing the drift of floating objects in real-world settings \citep{beron2019building,olascoaga2020observation,miron2020laboratory}.

\section{Balance between air and water drag} \label{sec:balance}

By contrast to previous leeway modeling studies, here we explicitly distinguish between two types of drag force, namely skin friction drag and form drag, both of which can be relevant for floating objects in typical geophysical settings. This is in part motivated by the use of separate drag terms for skin and form drag in iceberg models \citep[e.g.,][]{gladstone2001iceberg,martin2010parameterizing,marsh2015nemo}. Skin drag arises from the object's surface (its ``skin'') being subject to shear stresses as the object moves through a fluid. This friction effect is due to laminar or turbulent flow in the boundary layer close to the object's surface. Form drag, on the other hand, is determined by an object's size and shape. It is proportional to the cross-sectional area of the object normal to the direction of flow of the fluid. Both skin and form drag follow equivalent drag equations, scaling with the square of the relative velocity between fluid and object, but skin drag coefficients are typically three orders of magnitude smaller than form drag coefficients.

Membrane-like objects, such as oil slicks, are predominantly driven by skin drag. Form drag becomes dominant for objects with smaller length-to-height aspect ratios. Deeper keels will naturally give more importance to the role of water currents, whereas larger sails increase the sensitivity of an object's drift to wind forcing. Skin and form drag coefficients, density, and aspect ratio are thus central characteristics that will determine an object's sensitivity to winds and water currents. 

We note that in real-woarld settings the drag balance is complicated by the existence of two distinct types of turbulent boundary layers, both above and below the water surface: boundary layers associated with the air--water interface and boundary layers associated with the object--fluid interfaces. Identifying the appropriate free-stream fluid velocities for the drag equations in a setting with such complex boundary layer structures is nontrivial. For simplicity, here we assume that the dominant turbulent boundary layer above and below the floating object is that associated with the fluid--object interfaces. We thus ignore the role of the boundary layers associated with the air--water interface. The free-stream fluid velocities are referenced at a height that is greater than the combined height of the sail of the object plus the thickness of the object--air boundary layer, or below the depth of the keel plus the thickness of the object--water boundary layer. 

It may in general not be straight-forward to relate the free-stream wind velocity considered here to the commonly used 10-m surface wind velocity. For large objects such as icebergs with freeboard >10 m, the free-stream wind velocity should naturally be measured at some height above the freeboard. For objects with small freeboards, on the other hand, the relevant free-stream velocity should be considered at a height just outside the surface boundary layer -- potentially much lower than 10 m. While this free-stream velocity may in certain settings be notably different from the 10-m surface wind velocity, in many geophysical scenarios the two values are likely comparable, particularly when both values are measured outside the turbulent surface boundary layer.

We finally note that the exact structures of these turbulent boundary layers -- particularly the structure of the near-surface ocean layer -- remain a subject of ongoing research. 

Following \citet{martin2010parameterizing}, we write the drag force on the floating object due to water as:
\begin{equation}
\vec{F}_w = \frac{1}{2} (C^F_w A^V_w + C^S_w A^H_w) \rho_w |\Delta \vec{v}_w|\Delta \vec{v}_w, \label{F_eq}
\end{equation}
where $\Delta \vec{v_w} \equiv \vec{v}_w - \vec{v}$ denotes the velocity of the object $\vec{v}$ relative to the free-stream water velocity $\vec{v}_w$,  $C^F_w$ and $C^S_w$ are the form and skin drag coefficients for water, and $A^V_w$ and $A^H_w$ are the cross-sectional vertical and tangential horizontal surface areas of the object facing the relative flow of the water. Similarly, the drag force due to air is:
\begin{equation}
\vec{F}_a = \frac{1}{2} (C^F_a A^V_a + C^S_a A^H_a) \rho_a |\Delta \vec{v}_a| \Delta \vec{v}_a, \label{F_eq2}
\end{equation}
where the definitions of all terms are as given above for  \eqref{F_eq} except that the subscripts have been switched from $w$ for water to $a$ for air.

In geophysical settings, other forces including the Coriolis force may also play important roles in determining the drift of floating objects (as discussed in the introduction). The momentum equation for the drifting object can be written in the form:
\begin{equation}
m \d{\vec{v}}{t} = \vec{F}_w + \vec{F}_a  + \vec{F}_c + \vec{\mathcal{F}}, \label{mtm}
\end{equation}
where $m$ is the mass of the object, $\vec{F}_c=m \, f\, \kv \times \vec{v}$ is the Coriolis force with $f$ the Coriolis parameter and $\kv$ the vertical unit vector, and the term $\vec{\mathcal{F}}$ is a placeholder to represent any other forces acting on the object. The Maxey--Riley set mentioned above considers the force balance \eqref{mtm} for spherical particles that are immersed in unsteady and nonuniform flow \citep{beron2019building}.
 
In this study, we consider the limit where inertial effects, the Coriolis force, and other forces are small compared to air and water drag. In this case, the drift velocity is determined by a balance of drag forces due to near-surface water currents and winds. To provide a broad-strokes assessment of when this limit is applicable, we consider the ratio of the water drag force to the Coriolis force:
$$
\frac{|\vec{F}_w|}{|\vec{F}_c|} \sim \frac{\rho_w}{\rho}\frac{|\Delta \vec{v}_w|}{l \, f},
$$
where $\rho$ is the density of the object and $l$ its length. Here we have approximated that $C_w^F\sim 1$, $C_w^S=0$, $m = \rho A_w^V \, l$, and $|\vec{v}| \sim |\Delta \vec{v}_w|$, and we have neglected a factor of 2. For objects floating in the ocean, commonly $\rho_w/\rho \sim1$ and $|\Delta \vec{v}_w| \lesssim 0.1$ m/s. With a typical mid-latitude value for the Coriolis parameter of $f \sim 10^{-4}$ s$^{-1}$, we find that the drag force dominates the Coriolis effect when $l \lesssim 1$ km. Similarly, considering the ratio of air drag to Coriolis, we have:
$$
\frac{|\vec{F}_a|}{|\vec{F}_c|} \sim \frac{\rho_a}{\rho}\frac{|\Delta \vec{v}_a|^2}{l \, f \, |\vec{v}|^2}.
$$
In this case $\rho_a/\rho \sim10^{-3}$ and $\Delta v_a \sim 10$ m/s, resulting in dominant drag when $l \lesssim 10$ km.  In summary, the effects due to the rotation of the Earth are typically small when the size of the floating object is much less than $\mathcal{O}(1 \textrm{ km})$.
For larger objects, the Coriolis force becomes important. The force balance for drift of giant tabular icebergs, which is dominated by the Coriolis and pressure gradient forces, was previously considered in a similar framework by \cite{wagner2017jpo}.

Based on the discussion above, to determine the \emph{steady} drift of small floating objects -- with $l \ll \mathcal{O}(1 \textrm{ km}$) -- the air and water drag forces need to balance: 
\begin{equation}
 m \d{\vec{v}}{t} =  \vec{F}_w +\vec{F}_a  = 0. \label{balanced forces}
\end{equation}

Here, we consider idealized rectangular shapes such that:
$$
A^V_a = w \, b, \quad A^V_w = w \, d, \quad A^H_a = A^H_w = w \, l, 
$$
where $w$ is the object's width, $b$ its freeboard (height above the waterline), $d$ its draft (depth below the waterline), and $l$ its along-flow length. The across-flow width $w$ is the same for all surfaces and therefore cancels out in all calculations below. Henceforth, we only consider a two-dimensional framework, with vertical and along-flow dimensions. The freeboard and draft of the object can be calculated in terms of the object's height $h$, its density $\rho$, and the density of water $\rho_w$ as: 
\begin{equation}
b=h(1-\rho/\rho_w) \textrm{ and } d= h \rho/\rho_w.
\label{freeboard}
\end{equation}

In what follows, we limit our analysis to one dimension. In some settings, in particular for sea ice, it has been observed that the floating object typically drifts at a turning angle relative to the direction of the surface wind, due to the Coriolis force \citep{nansen1902oceanography,mcphee2008air}. For sea ice, this angle is between $0\deg$ and $40\deg$ to the right of the wind in the Northen Hemisphere \citep{lepparanta2011drift}. For oil slicks, \cite{stolzenbach1977review} report turning angles smaller than $10\deg$. In the cases of icebergs and mangrove drifters, turning angles are not often discussed; \cite{garrett1985} argue that this is due to the draft of icebergs being deep enough to not be significantly affected by the Ekman spiral. In the search-and-rescue literature, the deviation of an object's drift from the downwind direction is referred to as ``leeway divergence,'' which is not regarded to be primarily due to the Coriolis force, and ranges from $-30\deg$ to beyond $+30\deg$ \citep[see, for example, Figure 1 in][]{breivik2011wind}.
Here, for simplicity we focus on the drift speed in a one-dimensional framework, and we do not address the issue of turning angles. 

Since the air velocity, $v_a$, is typically much larger than the velocity of the object, $v$, we can approximate $\Delta v_a \equiv v_a - v \approx v_a$ \citep[e.g.,][]{stolzenbach1977review}. Substituting the drag expressions \eqref{F_eq} and \eqref{F_eq2} into the force balance \eqref{balanced forces} and solving for $v$, we find: 
\begin{equation}
  v = v_w + \gamma v_a, \label{gam1}
\end{equation}
where $\gamma$ is the leeway factor or wind factor and can be interpreted as the object's sensitivity to wind forces. Here, the wind factor is defined as:
\begin{equation}
    \gamma \equiv \sqrt{\frac{\rho_a}{\rho_w}}\sqrt{\frac{b C^F_a + l C^S_a}{d C^F_w + l C^S_w}}. \label{gamma}
\end{equation}
The four terms under the second square root represent the relative contributions of form and skin air drag (numerator) and form and skin water drag (denominator), respectively.  

Note that using the expressions \eqref{freeboard} for freeboard and draft, we could express $\gamma$ in \eqref{gamma} in terms of the four drag coefficients and three dimensionless parameters: the density ratio of air and water $\rho_a/\rho_w$, the length-to-height aspect ratio $l/h$, and the density of the object relative to that of water $\rho/\rho_w$. 

In the following we consider two limits, that of dominant skin drag (where $l \gg h$), and that of dominant form drag (where $l\ll h$). 

\subsection{The limit of dominant skin drag}

The limit where skin drag dominates over form drag, which we refer to as the ``membrane limit,'' has been investigated previously \citep[e.g.,][]{stolzenbach1977review}. In this limit, $l \gg h$ and thus the wind factor $\gamma$ simplifies to:
\begin{equation}
    \gamma \approx \sqrt{\frac{\rho_a}{\rho_w}}\sqrt{\frac{C^S_a}{C^S_w}}.
    \label{thin}
\end{equation}
If the skin drag coefficients are replaced by effective drag coefficients $C_a$ and $C_w$ that include form and skin drag, this expression for $\gamma$  becomes equivalent to the Nansen number, 
$N\!a \equiv \sqrt{\rho_a/\rho_w}\sqrt{C_a/C_w}$
\citep{lepparanta2011drift}.

For many typical materials the air and water skin drag coefficients are approximately equal, $C^S_a \approx C^S_w$ \citep{stolzenbach1977review}, in which case the wind factor \eqref{thin} reduces to $\gamma \approx \sqrt{\rho_a/\rho_w} \approx 3$\% (since $\rho_w \approx 10^3$\,kg m$^{-3}$ and $\rho_a \approx 1$\,kg m$^{-3}$). This explains why sufficiently thin objects typically drift at approximately 3\% of the wind speed relative to the water, regardless of the object's density and length-to-height ratio which do not appear in \eqref{thin}.  This membrane limit is indicated by the solid blue horizontal line in Figure~\ref{fig:gammavsrho}. For many floating objects the membrane limit is not appropriate, but we show below that a wide range of objects nonetheless have a wind factor that is similar to 3\%. 

In a recent study, \cite{samelson2020turbulent} showed that the flow velocity at the interface of two fluids, $v_{\rm int}$, is in general given by $|v_{\rm int} -v_2| = \frac{\sqrt{\rho_1}}{\sqrt{\rho_1}+\sqrt{\rho_2}}|v_1 - v_2|$, where $\rho_1$ is the density of the lighter fluid, $\rho_2$ the density of the heavier fluid, and $v_{1, 2}$ their corresponding free-stream velocities.  When $\rho_2 \gg \rho_1$ (such as is the case for air and water), the interface velocity reduces to $v_{\rm int} \approx v_2 +\sqrt{\rho_1/\rho_2}\, v_1$ \citep[Equation~(20) in][]{samelson2020turbulent}, which is identical to the expression for the membrane limit drift velocity above. We note that the approach of \cite{samelson2020turbulent} is different from the drag force balance we use here. Samelson's result arises from a condition of stress continuity at the fluid interface, while our result derives from balancing the drag forces acting on an object located at the fluid interface.  However, the stress continuity condition at the fluid interface (alongside a dimensional analysis argument that relates the surface stress to a velocity scale) implies that $\rho_a \tilde{v}^2_a = \rho_w \tilde{v}^2_w$, where $\tilde{v}_a$ and $\tilde{v}_w$ are the friction velocities for air and water \citep{samelson2020turbulent}. This condition is remarkably similar to the drag force balance in the membrane limit.  Note that a similar dimensional analysis argument is also used in deriving the drag force expressions~\eqref{F_eq} and~\eqref{F_eq2}. This explains the similarity between our wind factor expression in the membrane limit (and ignoring any lateral boundary layer effects associated with the finite horizontal extent of the floating membrane), given by our Equation \eqref{thin} and Equation~(20) in \citet{samelson2020turbulent}.

\subsection{The limit of dominant form drag}

Form drag coefficients are typically order 1, whereas skin drag coefficients are typically order $10^{-3}$. This implies that horizontal skin drag is the dominant force only when $l > 10^3h$. In cases where $l\ll10^3h$, which applies for example to most icebergs, mangrove drifters, and flotsam, the force balance on an object is predominantly a balance of the water and air form drags. Sea ice presents an intermediate case: while young, very thin, or uniformly grown sea ice (e.g., nilas) 
is predominantly influenced by skin drag, broken-up or deformed ice floes (e.g., featuring rafting or pressure ridges) are subject to substantial form drag \citep[e.g.,][]{arya1975drag,lu2011parameterization}.   

\begin{figure}[h]
\begin{center}
\includegraphics[width=\linewidth]{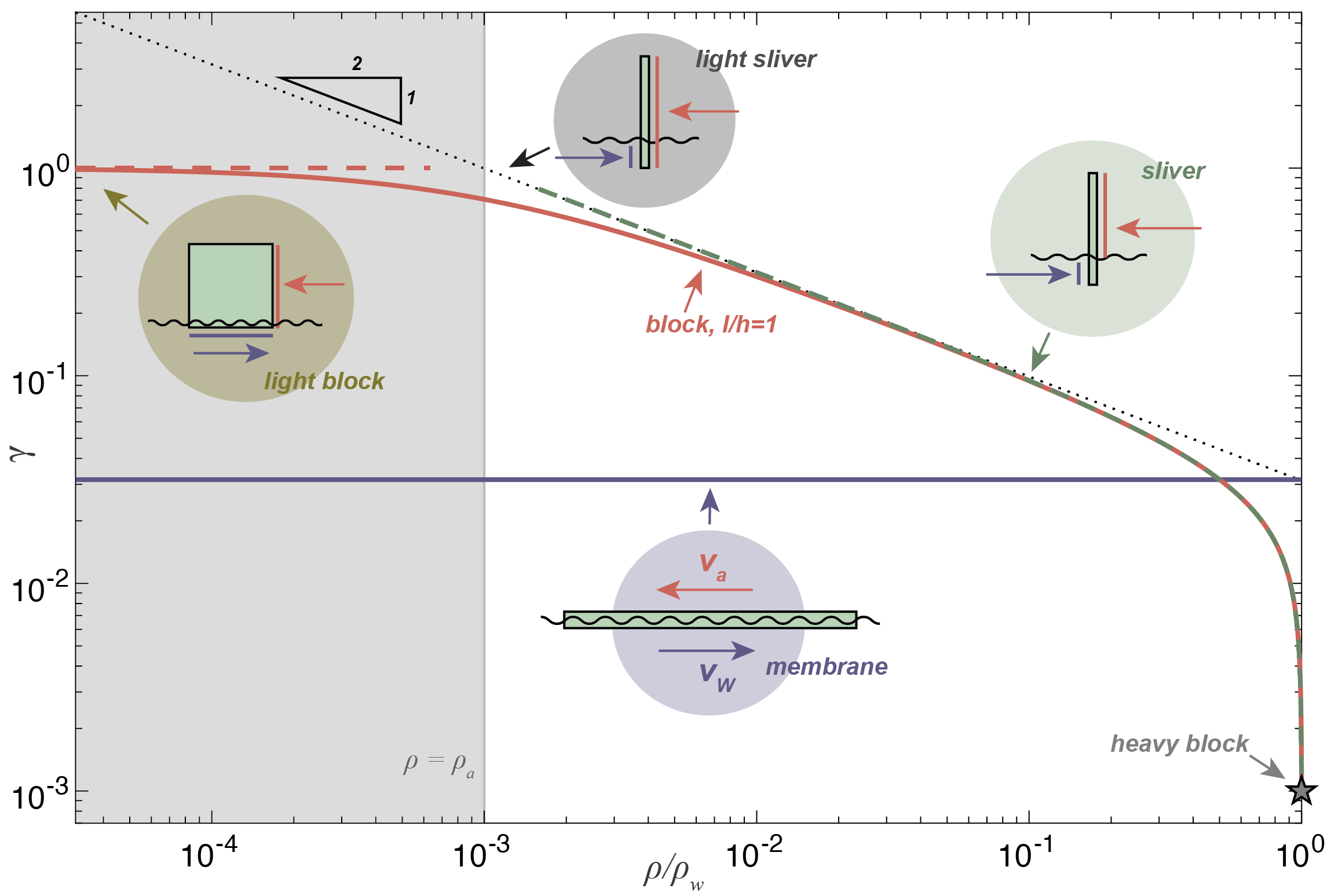}
\caption{Dependence of the wind factor, $\gamma$, on the density ratio $\rho/\rho_w$, for different aspect ratios $l/h$. Shown are the membrane limit ($l/h~\to~\infty$, purple horizontal line) 
and the case of a square block ($l/h = 1$, red solid line). The general sliver limit is shown ($l/h~\rightarrow~0$, green dashed), 
as well as the light sliver case where $\rho/~\rho_w~\rightarrow 0$ (dotted black line). In this limit $\gamma$ scales as $(\rho/\rho_w)^{-1/2}$, as indicated by the black triangle. The red dashed line indicates the limit of a light block where the force balance is between water skin drag and air form drag. 
Also shown is the limit of a the heavy square block (gray star). 
The insets indicate the dominant force balance in each case: red and blue arrows represent water and air velocity, and red and blue vertical and horizontal lines denote the surfaces on which the dominant drag components act. Note that for $\rho<\rho_a$ (gray shaded region), physical objects would float into the air.
}
\label{fig:gammavsrho}
\end{center}
\end{figure}

In the limit $l\rightarrow0$, the wind factor $\gamma$ reduces to:
\begin{equation}
    \gamma = \sqrt{\frac{\rho_a}{\rho_w}}\sqrt{\frac{b C^F_a}{d C^F_w}} = \sqrt{\frac{\rho_a}{\rho_w}}\sqrt{\frac{C^F_a}{C^F_w}}\sqrt{\frac{1-\rho/\rho_w}{\rho/\rho_w}}. \label{form}
\end{equation}
We refer here to an object in this limit as a vertical ``sliver,'' describing an (unphysical) object of finite height $h$ and zero along-flow length $l$. In Figure~\ref{fig:gammavsrho}, the sliver limit is indicated by a green dashed line, showing that this limit of \eqref{form} is essentially indistinguishable from the full solution of \eqref{gamma} (red solid line) for $l/h\leq1$ when $\rho/\rho_w>10^{-2}$.  The full solution shown in Figure~\ref{fig:gammavsrho} is that for a square block with $l/h=1$. For aspect ratios smaller than 1 this full solution will approximate the dashed sliver solution even more closely. Even for $l/h=10$, the sliver is still a good approximation as long as $\rho/\rho_w>10^{-1}$ (not shown). This encompasses most floating solid objects in geophysical settings. Since the air and water form drag coefficients are both order 1, and show up as a ratio under a square root (which brings the value closer to unity) in the expression for $\gamma$, we make the approximation $C^F_a = C^F_w$ in Figure~\ref{fig:gammavsrho}. Given the assumption of equal drag coefficients,  \eqref{form} is equivalent to Equation (16) in \cite{daniel2002drift}. 

Next, we consider two extremes for the object density: a light object where $\rho\ll\rho_w$, and a heavy object where 
$\rho \gg \rho_w$. If  $\rho\geq \rho_w $, the object is fully submerged or sinks, and the wind factor is $\gamma = 0$. More generally, as the object densities increase toward the limit $\rho/\rho_w \rightarrow1$, the wind factor scales as  $\gamma~=~\sqrt{\rho_a/\rho_w(1-\rho/\rho_w)}$ (not shown). 

Turning to the light object limit, if $\rho<\rho_a$ then the object would become airborne. As the object densities decrease toward the limit $\rho \to \rho_a$ the wind factor reduces to $\gamma = \sqrt{1-\rho_a/\rho_w} \approx 1$, since $\rho_a/\rho_w \approx 10^{-3} \ll 1$. This is in agreement with basic physical intuition: when an object's density approaches that of air, the object becomes no longer submerged in the water and its drift is fully determined by wind forcing (neglecting capillary effects and surface tension which are not included in this analysis). Considering instead the ratio of the density of the object and the density of water, in the light limit $\rho/\rho_w \ll 1$ we can approximate the last term in \eqref{form} as $(\rho/\rho_w)^{-1/2}$. The wind factor then reduces to $\gamma = \sqrt{\rho_a/\rho}$. This scaling is illustrated in Figure~\ref{fig:gammavsrho} (dotted line). 

Next, we consider the low density limit for the particular case of the square block with $l/h = 1$ (solid red line). 
Since the form drag coefficients are three orders of magnitude larger than skin drag coefficients, aspect ratios of order 1 lead to a force balance that is dominated by form drag for a large range of densities. 
Only when $\rho/\rho_w \lesssim 10^{-2}$ does the under-water skin drag become important and $\gamma$ diverges notably from the sliver approximation. In the limit $\rho/\rho_w \rightarrow 0$, the force balance is purely between the above-water form drag and the under-water skin drag, which gives $\gamma = \sqrt{\rho_a/\rho_w}\sqrt{C^F_a/C^S_w} \approx 1$ using $C^F_a/C^S_w \approx 10^3$ (red dashed line). For the physical lower bound $\rho = \rho_a$, we find $\gamma = 0.71$. That is to say, an idealized square block that is as light as air but still feels skin drag from the water will drift at 71\% of the wind velocity. Therefore, even in this extreme scenario the water skin drag still plays a substantial role in determining the object's drift. 

Finally, for a heavy block with $\rho = \rho_w$ traveling at the water surface, the force balance is between air skin drag and water form drag, giving $\gamma = \sqrt{\rho_a/\rho_w}\sqrt{C^S_a/C^F_w} \approx~10^{-3}$ (star marker in Figure~\ref{fig:gammavsrho}).

\section{Geophysical Parameter Range}

The higher the aspect ratio $l/h$ for an object, the less important the object's density becomes for determining the wind factor $\gamma$. In the membrane limit, $\gamma$ is independent of $\rho/\rho_w$ (Figure~\ref{fig:gammavsrho}), since the dominant force balance is between the above-water and below-water skin drags. The surfaces on which the skin stress acts do not change with density $\rho$, since the density only impacts how high in the water the object floats. On the other hand, for objects with small aspect ratios where form drags dominate the force balance, $\rho$ determines the relative above-water and under-water surface areas that the form stresses act on.

\begin{figure}[h!]
\begin{center}
\includegraphics[width=\linewidth]{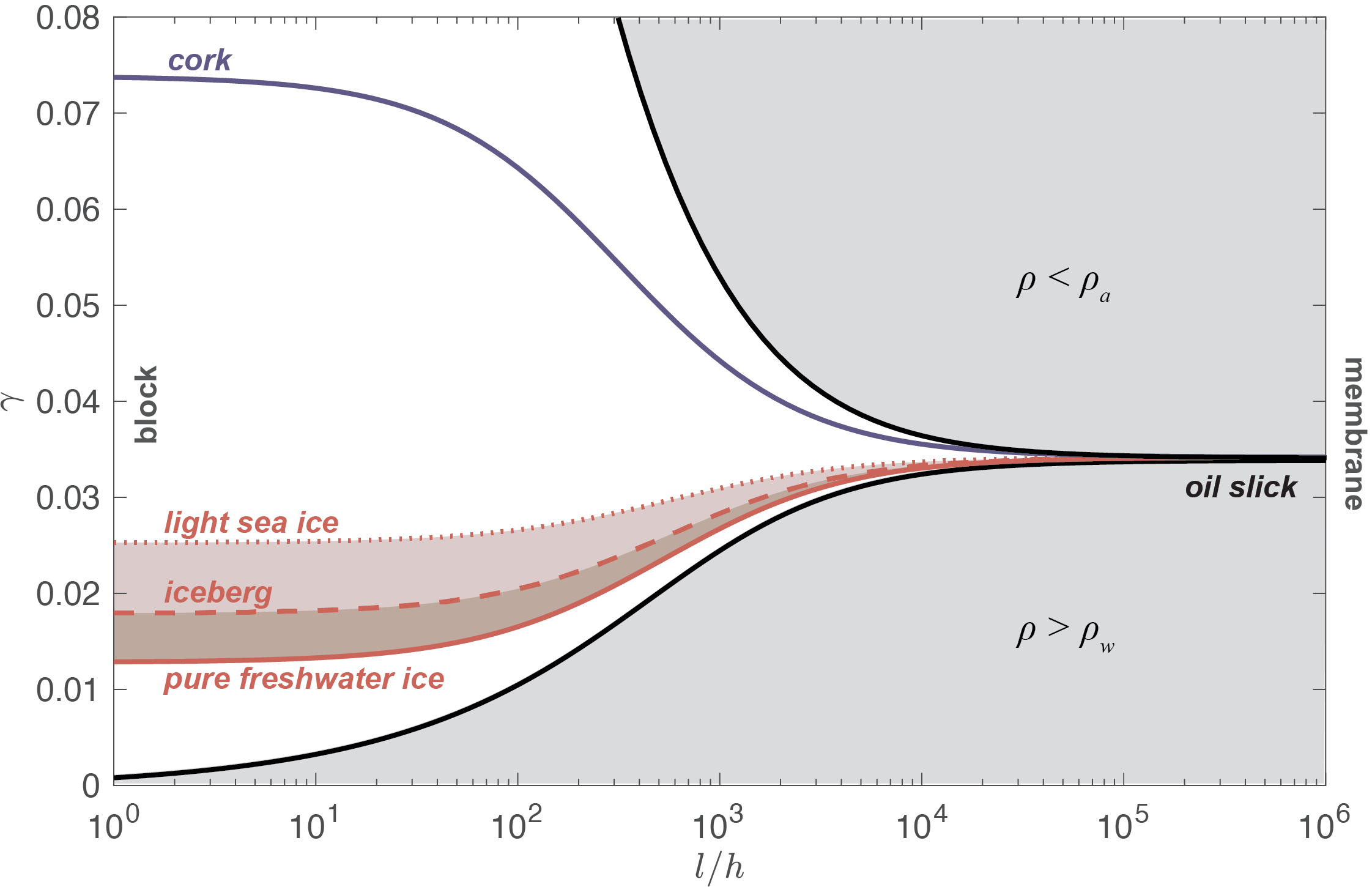}
\caption{The wind factor $\gamma$ as a function of the aspect ratio $l/h$, for a range of object densities $\rho$. In the membrane limit ($l/h\rightarrow \infty$), the wind factor reduces to $\gamma = \sqrt{\rho_a/\rho_w}=3.4$\%, independent of $\rho$. The lower bound for aspect ratios shown here is the square block, with $l/h=1$. In this case $\gamma$ is sensitive to density $\rho$. Values of $\gamma$ for ice of different densities (see text) are indicated, as well as the example of a material with the density of cork (250 kg m$^{-3}$). The gray shaded areas show the limits for which objects would start floating up into the air ($\rho<\rho_a$)
or sinking down into the water ($\rho>\rho_w$).}
\label{fig:gammavslh}
\end{center}
\end{figure}

Oil slicks and other thin membrane-like floating objects are therefore subject to a wind factor $\gamma \approx 3$\%, irrespective of their density. On the other hand, the wind factor of blocky icebergs and broken-up or deformed sea ice floes vary with the density of the ice. Measured sea ice densities are typically in the range of 740 kg m$^{-3}$ to 917 kg m$^{-3}$ \citep{timco1996}. Using \eqref{form} and assuming $C^F_a = C^F_w$, we find that this corresponds to an upper bound for the wind factor for sea ice $\gamma = 2.5$\%, and a lower bound $\gamma = 1.3$\% for pure ice (when $l/h = 1$). As $l/h$ increases, these upper and lower bounds converge to the membrane limit $\gamma = \sqrt{\rho_a/\rho_w}$ (Figure~\ref{fig:gammavslh}). An estimate of typical iceberg density, accounting for the presence of a firn layer, is $\rho = 850$ kg m$^{-3}$ \citep{bigg1997}, which gives  $\gamma = 1.8$\% (Figure~\ref{fig:gammavslh}). These theoretical values agree with observed ranges of $\gamma=1.6-1.8$\% for icebergs \citep{smith1983,garrett1985,bigg1997} and $\gamma=2.0-2.5$\% for sea ice \citep{nansen1902oceanography,zubov1945ice,browne1958,thorndike1982}. 

Naturally, objects that have small aspect ratios but are much lighter than sea ice will have substantially higher wind factors. For example, Figure~\ref{fig:gammavslh} illustrates the wind-factor dependence on $l/h$ for an object with the density of cork, $\rho = 250$ kg m$^{-3}$ (blue line). Here, the wind factor is $\gamma = 7.4$\% for $l/h = 1$. In this case, the freeboard is $b = 0.25h$ if the cork is floating in freshwater. Nevertheless, wind factors for objects investigated in search-and-rescue studies -- ranging from persons in water to unballasted life rafts and drifting fishing vessels -- are typically below 4\% \citep[Figure 1 in][]{breivik2011wind}.

\begin{figure}[ht]
\begin{center}
\includegraphics[width=.9\linewidth]{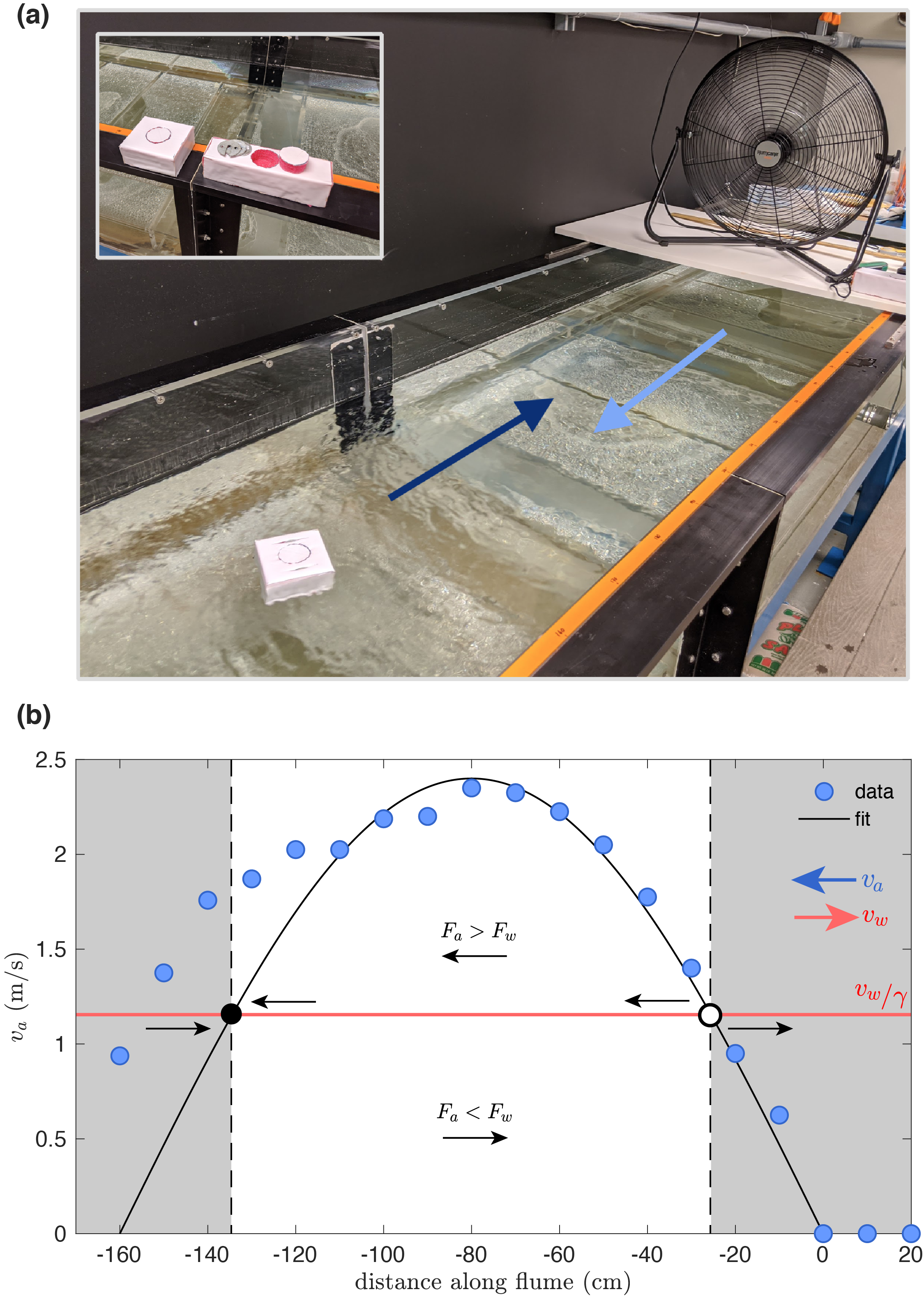}
\caption{(a) Experimental flume setup with fan and styrofoam block. The opposing directions of the water flow (dark blue arrow) and air flow (light blue arrow) are indicated. Inset: examples of two styrofoam blocks used, one of which has a few weights resting next to the cutout which has been temporarily removed for the weights to be inserted into the hole. (b) Measured mean air velocity at 10 cm above the water surface as a function of distance along the flume (blue circles). The black curve represents a visual best fit sine function, added purely for illustrative purposes. Air and water drag forces are balanced when the air speed is equal (and opposite) to $v_w/\gamma$ (red line, see text). In the gray shaded regions (where the black curve crosses below the red line) water drag is greater than air drag ($F_a<F_w$) and the object is advected to the right. When the black curve is above the red line drift occurs to the left, as indicated by the black arrows. The filled black circle therefore represents a stable equilibrium location, while the unfilled black circle represents an unstable equilibrium location.}
\label{fig:setup}
\end{center}
\end{figure}

\section{Flume Experiments}

We performed a set of experiments to assess the validity of the theoretical considerations above in a laboratory setting. The experimental setup (see Figure~\ref{fig:setup}a) consists of a custom-built 4-m horizontal flume and a 20-in household floor fan with adjustable speed located at the University of North Carolina Wilmington. The flume is filled with fresh water and has an adjustable water flow speed of up to 20 cm~s$^{-1}$. Beyond this upper limit significant ripples occur at the water surface. The flume speed is close to uniform for the upper 20 cm of the water column, and it can be set with an accuracy of about 1 cm~s$^{-1}$. The fan is positioned to blow in the opposite direction to the flume current. Styrofoam blocks of different geometries were sealed with impermeable tape, and we cut out openings for laboratory weights (Figure~\ref{fig:setup}a, inset).  Weights of different masses were added to vary the density of the blocks. 

The styrofoam blocks were then placed in the flume and would assume equilibrium resting positions at different distances from the fan where the drags due to the flume current and the fan wind balanced. 
From  \eqref{gam1}, we find that in this setting (for $\vec{v} = 0$) the wind factor is simply given by:
\begin{equation}
    \gamma = |{v_w}|/|{v_a}|.
\end{equation}
While the flume was set to produce a known flow speed $v_w$, the wind speed $v_a$ at the equilibrium location of the styrofoam block was measured by hand, using a Kestrel 1000 wind anemometer at a height $z = 10$ cm above the water surface and taking 10-second average values. These readings were not sensitive to wind measurement height in the range $z \approx 1 - 20$ cm above the water surface. This method was made possible by the air speed varying considerably with distance from the fan: it would increase over a span of approximately 80 cm from 0 to almost 2.5 m~s$^{-1}$ (for the fan at its lowest setting), and then decrease gradually with increasing distance from the fan (see Figure~\ref{fig:setup}b). While the figure shows that the profile of increasing air velocities presented good agreement with a sinusoidal fit, the decreasing profile was less regular, and it was notably impacted by the wall near the end of the flume.  The more reliable readings were therefore obtained by determining the location of the unstable equilibrium point closer to the fan (unfilled black circle in  Figure~\ref{fig:setup}b), rather than the stable equilibrium point further from the fan (solid black circle). This was done by placing the styrofoam blocks repeatedly in the flume and observing whether they would eventually be advected by the water current to the right (when $F_a<F_w$), or by the wind to the left (when $F_a>F_w$). An approximate equilibrium location was readily found using this scheme.

We note that more sophisticated experimental setups, such as that of \cite{miron2020laboratory}, use a water flume--wind tunnel setup that features constant air and water speeds throughout. In such cases, the force balance is determined from the (approximately) constant velocity of the floating object, rather than considering a cancelling of the drag forces as we have done here.

The styrofoam blocks had a thickness of $h = 5$ cm, lengths $l = 5$ cm or $l = 10$ cm, and widths $w = 5$, 10, or 20 cm. Here $l$ is the along-flow dimension of the block, and $w$ is the across-flow dimension. The different values of $w$ were used to test the assumption that the across-flow dimension does not significantly impact the wind factor in this setting. Adding laboratory weights to the blocks allowed us to vary the densities in the range $0.03 < \rho/\rho_w < 0.91$.

\begin{figure}[h!]
\begin{center}
\includegraphics[width=\linewidth]{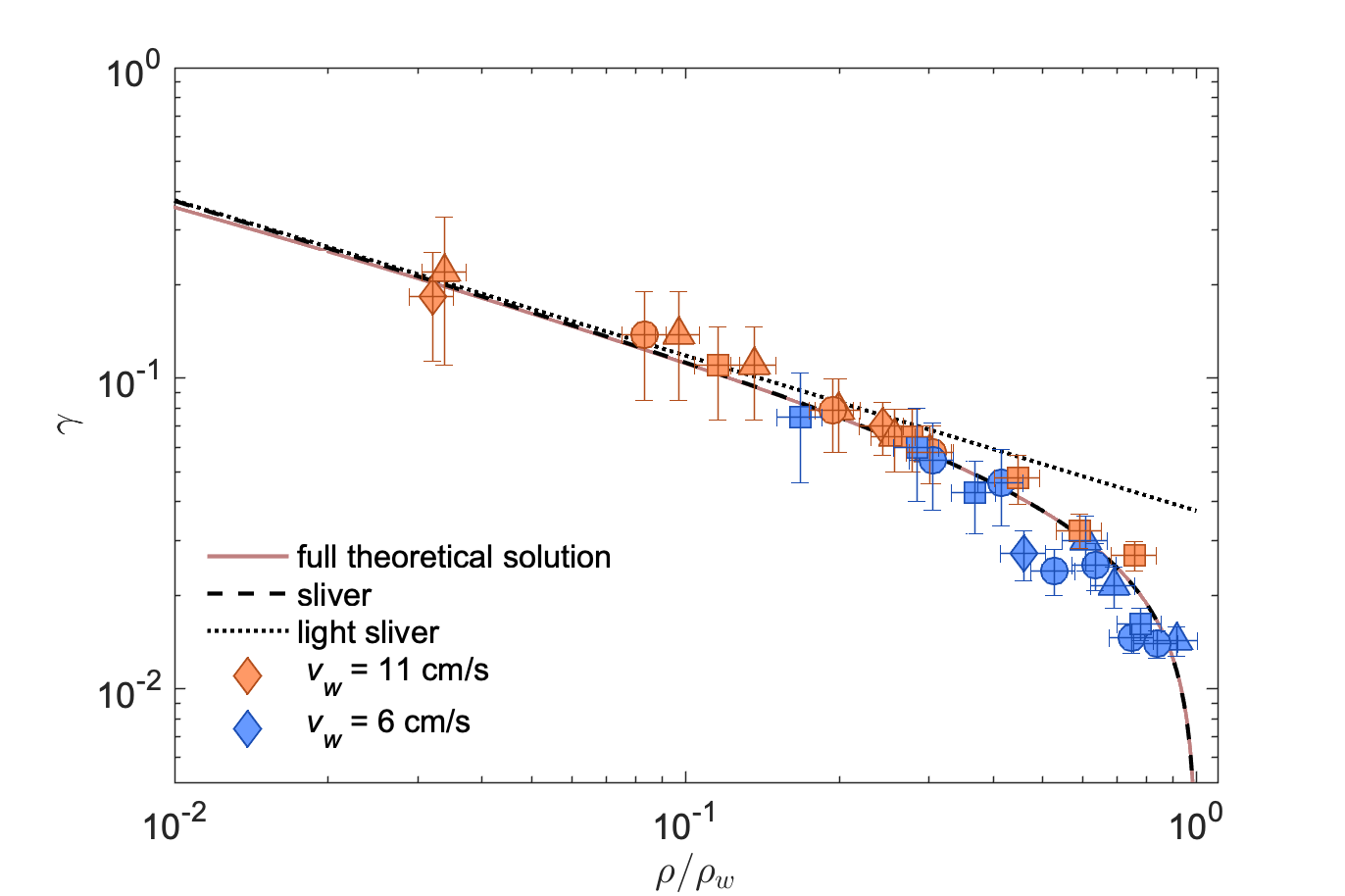}
\caption{Dependence of wind factor on density for flume experiments. Red and blue symbols represent experiments with the flume set to different water speeds as indicated in the figure legend. The symbols correspond to different object geometries: $l/h$ is 1 ($\diamond,\Box$) or 2 ($ \circ, \triangle$). The across-drag dimension of the blocks $w/l$ is 1 ($\Box,\circ$), 2 ($\diamond$), or 4 ($\triangle$). 
The theoretical lines are as in Figure~\ref{fig:gammavsrho}, using the best fit ratio of form drag coefficients, $C^F_a/C^F_w = 1.4$.}
\label{fig:expt1}
\end{center}
\end{figure}

Since the shapes of the styrofoam blocks were restricted to aspect ratios $l/h$ of 1 and 2, these experiments are only suitable to test the validity of the theory in the sliver limit (where form drag dominates). The experimental results show that the wind factor was approximately insensitive to which aspect ratio was used (Figure~\ref{fig:expt1}), consistent with the theoretical results for the sliver limit. Similarly, no discernible dependence of $\gamma$ on the across-flow dimension $w$ was observed. Two flume settings were used: a faster flow of 11 cm s$^{-1}$ mostly for lighter blocks (since these were more sensitive to the fan's wind forcing), and a slower flow of  6 cm s$^{-1}$ for heavier blocks (otherwise the fan's peak wind forcing was not sufficient to balance the water drag). The observed wind speeds at the location of stationary styrofoam blocks fell in the range $0.5 < v_a < 4.3$ m s$^{-1}$. This gives a span of wind factors of $\gamma = 1.4$\% for the densest block to $\gamma = 22$\% for the lightest block.

For $\rho/\rho_w > 0.4$, the observed wind factors deviate notably from the theoretical ``light sliver'' limit and fall closely on the curve of the general sliver limit. This highlights that in this limit the force balance truly is between the below-water and above-water form drags. 

In Section \ref{sec:balance}, we approximated that $C^F_a/C^F_w = 1$. Previous work on drag coefficients for icebergs \citep{kubat2005,keghouche2009} and other objects have given values for this ratio, which tends to be similar to 1. Here, we can use the observed wind factor dependence, $\gamma(\rho/\rho_w)$, to estimate the ratio $C^F_a/C^F_w$ for the blocks in the flume experiments. This is done by varying $C^F_a/C^F_w$ in the theoretical estimate for $\gamma$ and computing the root-mean-square error between the resulting theoretical curve and the observed values of $\gamma$. We find that for our experimental configuration, the root-mean-square error is minimized when we take the ratio to be $C^F_a/C^F_w = 1.4$. For this drag ratio, we observe strikingly close correspondence between the observed and theoretical estimates for the full range of block densities in the experiments (Figure~\ref{fig:expt1}). 
Notably, this ratio for the styrofoam blocks is in close agreement with the findings for icebergs of \cite{bigg1996}, who derive this ratio from iceberg trajectories and report $C^F_a/C^F_w = 1.44$.   

The experimental setup imposes several limitations on the parameter range that can feasibly be explored: (i) The wind speeds have an upper limit of $\sim 4.5$ m/s and the flume speed has a lower limit of $\sim 5$ cm/s. According to the theory, this gives a lower limit for the experimental wind factor of $\gamma \sim 1.1$\%. This also implies an upper limit for the object density of $\rho/\rho_w = 0.91$ (which is roughly that of pure ice). For densities greater than that, the maximum fan strength would not be sufficient to balance the water drag, even at the lowest flume setting. (ii) To explore the low-density regime, styrofoam blocks were hollowed out such that the effective density was as low as $ \rho/\rho_w = 0.03$. It was found not to be experimentally feasible to find steady state positions for blocks with densities lower than that. (iii) To obtain observable deviations from the sliver limit (i.e., where skin drag becomes important), we estimate that the aspect ratio has to be of the order $l/h \gtrsim 100$ (see Figure~\ref{fig:gammavsrho}). The experimental setup was not suitable for sheets of such small thickness: surface ripples and capillary waves did not allow for top surfaces that were consistently splash free or bottom surfaces that did not feature bubbles, thus distorting the drag balance considerably. Hence we were not able to approach the membrane limit in the flume experiments.

\subsection{Previous Flume Experiments With Mangrove Propagules}

\begin{figure}[h!]
\begin{center}
\includegraphics[width=\linewidth]{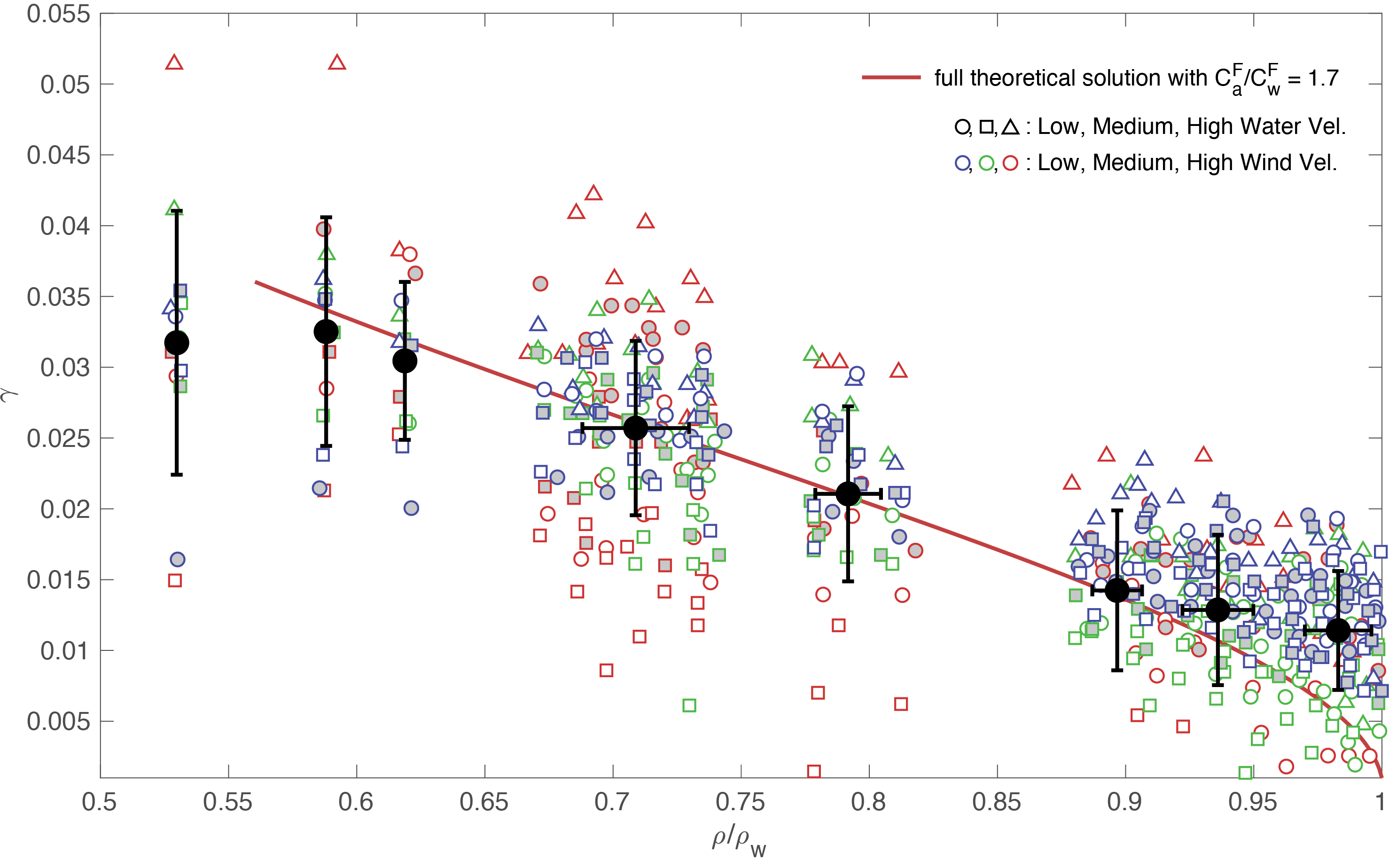}
\caption{Previous flume experiment results derived from data digitized from \cite{vanderstocken2015}. Each black marker corresponds to one species of mangrove. Error bars are $\pm 1 \sigma$ for all measurements for a given species. The different colors of the small markers correspond to different wind strengths in the flume experiments: Low 2.7 m/s (blue), Medium 4.5 m/s (green), and High 6.0 m/s (blue). The different markers correspond to different water velocities: Low 0 m/s ($\bigcirc$), Medium 0.15 m/s ($\square$), and High 0.3 m/s ($\triangle$). Gray-filled markers denote experiments where the wind direction was opposite the water flow; white-filled markers denote experiments where it was aligned with the water flow. For most species,  the theoretical solution with $C^F_a/C^F_w = 1.7$ (solid line) yields a relatively close fit.}
\label{fig:expt2}
\end{center}
\end{figure}

\cite{vanderstocken2015} report on flume experiments with a setup similar to the one presented above, investigating the drift of different species of mangrove propagules. The authors carried out a set of 16 different experiments for each species with either no wind or one of three different wind velocities (Low, 2.7 m/s; Medium, 4.5 m/s; High, 6.0 m/s), three different water velocities (Low, 0 m/s; Medium 0.15 m/s; High, 0.3 m/s), and two wind directions (along the water flow and against the water flow). The densities of the propagules fall in the range $\rho/\rho_w = 0.5 - 1$. The shapes for the different species vary widely, from the relatively light \textit{H. littoralis} whose propagules resemble ``small sailboats'' to the dense spherical \textit{X. granatum} which is known as the ``cannonball'' \citep{vanderstocken2015}. 

We compute the wind factor from the reported velocities as $\gamma(\rho) = [v(\rho)-v_w]/v_a$, where $v(\rho)$ was digitized from Figure 2 in \cite{vanderstocken2015}. One may have expected that $\gamma$ would extend over a large range for this widely varied set of experiments. However, even for the lightest propagules we find $\gamma < 5$\% (Figure~\ref{fig:expt2}). For most of the propagules lighter than $\rho/\rho_w = 0.85$ we find $\gamma = 1.5 - 3.5$\%, while for $\rho/\rho_w > 0.85$ we typically find $\gamma = 0.5 - 2.5$\%. The mangrove propagules are thus subject to wind factors that are broadly similar to those for ice and oil. 

One might further expect a large spread in the ratio of drag coefficients due to the very different shapes of the propagules. Yet, for most species the best fit to the theoretically computed $\gamma$ is obtained using $C^F_a/C^F_w \approx 1.7$ (Figure~\ref{fig:expt2}).
Outliers are the lightest propagules and the densest ones. 
We note that for the densest species, the surface area and freeboard exposed to the wind becomes very small, on the order of the capillary length of water ($\sim 3$ mm), which means that effects such as surface tension and wind-induced ripples in the water surface are expected to play a non-negligible role. The length-to-height aspect ratios range from vertically drifting, sliver-like \textit{C. tagal} with $l/h\ll1$ to thin, flat  \textit{R. mucronata} with $l/h\approx30$ \cite[see][their Figure S1A]{vanderstocken2015}. All of these fall nevertheless near the limit of dominant form drag in the theory presented above.

\section{Conclusions}

We found that in many geophysical settings, the drift of floating objects is predominantly determined by a balance of near-surface winds and water currents, and the sensitivity to wind forcing falls in the relatively narrow range of $\gamma \simeq 3 \pm 2$\%. We explained the behavior using an analytical solution of an approximate force balance and showed that this solution holds for sea ice, icebergs, oil slicks, mangrove propagules, and a wide range of man-made floating objects. The behavior is similar despite the widely varying densities, shapes, and sizes of these floating objects. We showed that this dependence essentially occurs due to the wind sensitivity being approximately equal to $\sqrt{\rho_a/\rho_w} \approx 3$\%. The different characteristics of these different objects typically only result in a small correction to this number.  The drift of most objects is set by a balance of form drags, with skin drag playing a negligible role, except for the membrane-like oil slicks and to a certain degree sea ice. The flume experiments presented here, as well as those with mangrove drifters by \cite{vanderstocken2015}, can serve two practical purposes: \emph{(i)} a straight-forward way to establish the wind sensitivity of a given type of floating object and \emph{(ii)} a low-tech approach to estimate the form drag ratio $C^F_a/C^F_w$, under the assumptions of these theoretical calculations.  The theoretical considerations presented here provide a physically intuitive framework to parse out the relative importance of water currents and winds in determining the drift of floating objects.

\acknowledgments
This study was supported by NSF OPP grants 1643445 and 1744835. N.C.C.~was supported by the Australian Research Council DECRA Fellowship no.~DE210100749. Without implying their endorsement, we thank Emma Beer, Roger Samelson, Juan Restrepo, and Tapio Schneider for helpful conversations about this project.

\datastatement
The experimental data is available at https://github.com/tillwagner/drift.

\end{document}